\documentclass[twocolumn,prl,superscriptaddress,preprintnumbers]{revtex4}[11pt]
\pdfoutput=1
\usepackage{amsmath,amssymb,graphicx} 
\usepackage{epsf,verbatim}
\usepackage{hyperref}
\usepackage{comment}
\usepackage{color}

\usepackage{subfigure}

\def\simgt{\mathrel{\lower2.5pt\vbox{\lineskip=0pt\baselineskip=0pt
           \hbox{$>$}\hbox{$\sim$}}}}
\def\simlt{\mathrel{\lower2.5pt\vbox{\lineskip=0pt\baselineskip=0pt
           \hbox{$<$}\hbox{$\sim$}}}}

\newcommand{\mysubsection}[1]{{\bf{#1}.}}

\newcommand{\abs}[1]{\left| #1 \right|}
\newcommand{\unit}[1]{\mathrm{\ #1}}

\newcommand{\squishlist}{
 \begin{list}{$\bullet$}
  { \setlength{\itemsep}{0pt}
     \setlength{\parsep}{3pt}
     \setlength{\topsep}{3pt}
     \setlength{\partopsep}{0pt}
     \setlength{\leftmargin}{1.5em}
     \setlength{\labelwidth}{1em}
     \setlength{\labelsep}{0.5em} } }

\newcommand{\squishlisttwo}{
 \begin{list}{$\bullet$}
  { \setlength{\itemsep}{0pt}
     \setlength{\parsep}{0pt}
    \setlength{\topsep}{0pt}
    \setlength{\partopsep}{0pt}
    \setlength{\leftmargin}{2em}
    \setlength{\labelwidth}{1.5em}
    \setlength{\labelsep}{0.5em} } }

\newcommand{\squishend}{
  \end{list}  }

\newcommand{\be}{\begin{equation}}
\newcommand{\ee}{\end{equation}}
\newcommand{\bea}{\begin{eqnarray}}
\newcommand{\eea}{\end{eqnarray}}
\newcommand{\OO}{\mathcal{O}}

\newcommand{\f}{\frac}
\renewcommand{\k}{{\bf k}}
\newcommand{\x}{{\bf x}}
\newcommand{\G}{{\bf G}}
\newcommand{\q}{{\bf q}}
\renewcommand{\v}{{\bf v}}

\newcommand{\qtyp}{q_0}
\newcommand{\aprime}{A_D}

\def\DM{_{\rm DM}}

\newcommand{\xenon}{{\sc Xenon10}}

\begin{document}

\title{Direct Detection of Sub-GeV Dark Matter}%

\author{Rouven Essig}
\affiliation{SLAC National Accelerator Laboratory, Stanford University, Menlo Park, CA 94025}

\author{Jeremy Mardon}
\affiliation{Berkeley Center for Theoretical Physics, Department of Physics, University of California, Berkeley, CA 94720}
\affiliation{Lawrence Berkeley National Laboratory, Berkeley, CA 94720}
\affiliation{Stanford Institute for Theoretical Physics, Department of Physics, Stanford University, Stanford, CA 94305}

 \preprint{SLAC-PUB-14538}
 
\author{Tomer Volansky}
\affiliation{Berkeley Center for Theoretical Physics, Department of Physics, University of California, Berkeley, CA 94720}
\affiliation{Lawrence Berkeley National Laboratory, Berkeley, CA 94720}

\begin{abstract}
\noindent 
Direct detection strategies are proposed for dark matter particles with MeV to GeV mass. In this largely unexplored mass range, dark matter scattering with electrons can cause single-electron ionization signals, which are detectable with current technology. Ultraviolet photons, individual ions, and heat are interesting alternative signals.  Focusing on ionization, we calculate the expected dark matter scattering rates and estimate the sensitivity of possible experiments.  Backgrounds that may be relevant are discussed.  Theoretically interesting models can be probed with existing technologies, and may even be within reach using ongoing direct detection experiments.  Significant improvements in sensitivity should be possible with dedicated experiments, opening up a window to new regions in dark matter parameter space.

\end{abstract}

\maketitle

 \setcounter{equation}{0} \setcounter{footnote}{0}

\section{INTRODUCTION}
\label{sec:intro}
The identity of Dark Matter (DM) is unknown.  
The well studied paradigm of DM consisting of Weakly Interacting 
Massive Particles (WIMPs) with masses around the Weak scale is 
attractive: a WIMP naturally has the correct thermal relic 
abundance and appears in many new physics models that explain the 
hierarchy problem. 
A WIMP is also an ideal experimental target, with many 
direct and indirect DM and collider experiments currently searching for it. 
It is possible, however, that this theoretical prejudice  has been misleading. 
In particular, despite significant experimental effort, no unambiguous direct or indirect 
evidence for WIMPs has been obtained to date. 
It is important therefore to explore other theoretically motivated scenarios.

An interesting possibility is {\em light DM (LDM), with masses in the keV to GeV range}.  Such LDM is theoretically motivated and may naturally occur if DM does not couple strongly to the visible sector. In particular, the mass of a particle residing in a hidden sector may originate from Weak scale dynamics but be suppressed by small couplings between the hidden and visible 
sectors (see e.g.~\cite{ArkaniHamed:2008qn,Pospelov:2008jd,ArkaniHamed:2008qp,Cheung:2009qd,Essig:2010ye} and references therein).  
While considerable study is still in order, many existing models can accommodate LDM, including 
WIMPless~\cite{Feng:2008ya}, ``MeV''~\cite{Boehm:2003hm,Boehm:2003ha,Borodatchenkova:2005ct,Pospelov:2007mp,Fayet:2007ua,Hooper:2008im}, asymmetric~\cite{Nussinov:1985xr, Kaplan:1991ah, Kaplan:2009ag, Falkowski:2011xh}, bosonic super-WIMP~\cite{Pospelov:2008jk}, 
 Axino~\cite{Rajagopal:1990yx,Covi:1999ty,Choi:2011yf}, 
 gravitino \cite{Ellis:1984eq}, and sterile neutrino DM~(see review in \cite{Kusenko:2009up}).

In this letter, we focus on the MeV to GeV mass range.  We argue that 
simple experimental setups can allow for the direct detection of LDM and 
can probe a wide class of models. 
The ability to detect the signals of LDM scattering is already within reach with existing technologies, 
as will be demonstrated explicitly with XENON10 data in \cite{XE10}, and might also be possible with 
current direct detection experiments such as 
XENON100~\cite{Aprile:2011dd}, LUX~\cite{Akerib:2011ix}, and CDMS~\cite{Akerib:2005zy}.  
Dedicated experiments may significantly improve the sensitivity for LDM.  This letter 
aims in part at initiating the effort towards probing this mass range with direct detection experiments. 
A more comprehensive discussion of possible direct detection avenues is postponed to future work.

\section{BASIC PROPOSAL}
\label{sec:proposal}
Current direct detection experiments search for nuclear recoils caused by DM scattering.   For LDM, the average energy transferred in an elastic nuclear recoil is $E_{\rm nr} = q^2/2m_N \!\simeq\! 1\unit{eV}\!\times (m\DM/100\,\text{MeV})^2 (10\,\text{GeV}/m_N)$, 
where $m_N$ is the mass of the nucleus, $q\!\sim\! m\DM v$ is the momentum transferred, and $v\!\simeq\!10^{-3}$ is the DM velocity. This {\em nuclear} recoil energy is
well below the lowest thresholds achieved in existing direct detection experiments.  Consequently, vanilla elastic scattering with the nucleus does not allow for the detection of DM much below the GeV mass scale. 

In contrast, the total energy available in the scattering is significantly larger, $E_{\rm tot} \simeq  m\DM v^2/2\!\simeq\!50\,\text{eV}\!\times\! (m\DM/100\,\text{MeV})$, and 
may easily suffice to trigger  inelastic atomic processes that could lead to visible signals. 
We identify three leading possibilities:
\begin{itemize}
\item \textit{Electron ionization (DM--electron scattering).}
\item \textit{Electronic excitation (DM--electron scattering).}
\item \textit{Molecular dissociation (DM--nuclear scattering).}
\end{itemize}
These processes typically require energies of 1--10 eV, and so may be caused by scattering of DM particles with mass as small as $\OO$(MeV), through interaction with electrons, nuclei, or the electromagnetic field (e.g.~via higher dimension operators).  The resulting signals are small, but the technology to detect them is feasible, and in some cases already established. Three types of signals that may be particularly promising are~\footnote{
Photons~\cite{Starkman:1995ye} and ionization~\cite{Bernabei:2007gr, Dedes:2009bk, Kopp:2009et} have previously been considered as signals of DM--electron scattering, explicitly in the context of WIMPs. In both cases the rates are extremely small; indeed one of the main results of~\cite{Kopp:2009et} is that for WIMPs, nuclear recoil signals always dominate over electron scattering signals, (almost) model-independently. Only for LDM, for which nuclear recoil signals are unobservable, does electron scattering become important (in fact crucial) for direct detection.
}:
\begin{description}
\item \textit{Individual electrons.} 
An electron may be ionized (or, in semiconductors, excited to a conduction band) by DM--electron scattering. Signal amplification can be achieved in certain materials by drifting the electron in an applied electric field, causing it to scatter and produce an observable secondary signal. 
The primary recoiling electron can also ionize other electrons.  
\item \textit{Individual photons.} Following an inelastic process such as atomic excitation, de-excitation may produce photons, which could escape the target and be detected if they are not efficiently reabsorbed.  This may require a two (or more) step de-excitation, in which at least one photon does not sit on a resonance for reabsorbtion and can propagate over long distances~\cite{Starkman:1995ye}. Such multi-step de-excitations are natural in atoms and molecules. The main experimental challenge for detecting individual photons lies in reducing the noise and dark count levels.  Current capabilities seem to imply a somewhat higher (but still potentially interesting) experimental threshold, as only signals with more than one photon would be resolved above noise.
\item \textit{Individual ions.}  Ions could be produced either by ionizing electrons, or as the result of molecular dissociation.  The latter probes primarily nuclear rather than electronic interactions, and so may be an interesting complimentary direction to pursue. The technology, however, for using molecular targets and detecting individual ions still needs to be established.
\item \textit{Heat/phonons.} 
Much of the energy deposited by LDM scattering may emerge as 
phonons or heat, especially if any charge carriers produced are not drifted away from the interaction site by an electic field. This may be detectable with ultra-low threshold bolometers, such as the one recently proposed in~\cite{Formaggio:2011jt}.
\end{description}

A discovery of DM may be possible by searching for one or more of the above signals.  
Since the backgrounds to these signals are currently not well understood (see below), 
it remains to be seen whether the background discrimination capabilities found in current WIMP 
searches can be achieved.  In any case, a discovery is possible through the observation of the annual 
modulation of the signal~\cite{Drukier:1986tm}.
To illustrate the principle of LDM direct detection, we focus for the remainder of this letter 
on the detection of individual electrons produced by DM-electron scattering. 
We postpone further study of the prospects for LDM searches using photons, phonons, and ions to future work.

The capability to measure single electrons was demonstrated in both the ZEPLIN-II~\cite{Edwards:2007nj} and \xenon~\cite{Sorensen:thesis, Angle:2011th} experiments. This depends in both cases on the physical amplification achieved by drifting the electrons through gas-phase xenon, which produces detectable scintillation photons. The same principle works in semiconductor targets, where drifting electrons induce observable phonons. Low threshold detectors may be achieved by maximally exploiting this effect~\cite{Neganov:2001bn}, including the ``CDMS Light'' mode of operation~\cite{Tali:private-comm} of CDMS ZIP detectors.   We note that, while lacking single electron detection capability, CDMS Light may be able to probe LDM, calling for a careful study.  
A further possibility is to apply a large drift field where a single electron can trigger an avalanche, leading to a potentially observable current. This has been demonstrated in gas based detectors~\cite{Gorodetzky:1999ty}, and proposed in semiconductor detectors~\cite{Starostin:1999jj}. 
In summary, a variety of detection principles sensitive to LDM scattering seem realistic.

\section{DIRECT DETECTION RATES}
\label{sec:rates}

We now present formulae for the rates of LDM scattering in a target material to produce observable electrons~
\footnote{Our calculation differs from others in the literature~\cite{Bernabei:2007gr, Dedes:2009bk, Kopp:2009et} in several important respects.  
In~\cite{Bernabei:2007gr}, the electron inside the atom is incorrectly treated as being effectively free.
In~\cite{Dedes:2009bk, Kopp:2009et}, the binding of the target electron is correctly treated, but the effect of the binding potential on the \emph{recoiling} electron is dropped.
We include this effect, which can enhance the scattering rate by as much as two orders of magnitude for the most common recoil energies (a few 10's of eV).
(The effect is not important for the electron recoil energies considered in~\cite{Kopp:2009et}, of a few keV).
In addition,~\cite{Dedes:2009bk} considers only hydrogenic electron orbitals, and~\cite{Kopp:2009et} only spherically symmetric inner shells. We also consider, for the first time,  ionization (more precisely, excitation of valence electrons to conduction bands) in semi-conductor crystals, accounting for the full band structure.}. 
For atomic or molecular materials this means ionizing an electron, while for semiconductors (and insulators) it means exciting a valence electron to a conduction band. The cross sections for these processes involve atomic form-factors and may significantly differ from scattering with a free electron.  Indeed, the presence of the binding potential introduces two competing effects, one which acts to enhance the scattering cross section and the other to suppress it.   

The enhancement occurs due to the attractive potential around the nucleus. Semiclassically, energy conservation implies that an electron that escapes with momentum $p$ far from the atom must have initially scattered with some larger momentum $p_0$. The volume of phase space available is then $p_0^2 d p_0$, rather than the smaller $p^2 d p$, and the scattering rate is increased correspondingly. More formally, the effect is due to the distortion of the escaping electron wavefunction in the vicinity of the atom. It is familiar from beta decays, where the differential rate is enhanced by the Fermi-factor $F(p,Z_{\rm eff})=\abs{\psi_{\rm exact}(0)/\psi_{\rm free}(0)}^2$. In the non-relativistic limit it takes the form 
\begin{equation}
\label{eq:fermifunc}
F(p,Z_{\rm eff}) = \frac{2\pi\eta}{1-e^{-2\pi\eta}} \,,\qquad \eta = Z_{\rm eff} \frac{\alpha m_e}{p}\,,
\end{equation}
where $Z_{\rm eff}$ is the effecive charge felt by the escaping electron.
$F(p,Z_{\rm eff})$ grows as $1/p$, as a slowly-escaping electron is more affected by the potential well.   
This is nothing other than the Sommerfeld enhancement (for a concise review see~\cite{ArkaniHamed:2008qn}), but occurring to an outgoing rather than an incoming state. For the case at hand the interaction is delocalized across the atom, whereas in beta decay it is confined to the origin. However the effect is qualitatively the same, and in both cases the low-$p$ behaviour is straightforward to derive from the phase space argument.

Due to the uncertainty in its initial momentum, a bound electron may escape with a given momentum $p$ after recieving \emph{any} momentum transfer $q$. However, there is a significant penalty on those regions of phase space where $q$ deviates too far from the typical size, $\qtyp$, associated with the atomic process. This can come into conflict with the kinematic requirement on the DM velocity needed to overcome the electron's binding energy,
\begin{equation}
\label{eq:vmin}
v > v_{min} = \frac{\Delta E_B + E_R}{q} +\frac{q}{2 m_\chi} \, ,
\end{equation}
where $\Delta E_B + E_R$ is the total energy transferred to the electron (binding $+$ recoil). 
Given that the typical size of the DM velocity is $10^{-3}c$, transitions in which $\Delta E_B \simgt 10^{-3} \qtyp$ receive a suppression relative to free electron scattering.
It follows that one way to maximize rates is to use elements with high $Z$, which exhibit a deep potential, while another is to minimize $\Delta E_e$ by using semiconductors targets.  Since electrons then only need to be excited across the band gap, the energy required is significantly smaller.

We now assume DM interacts directly with electrons, and parametrize its coupling in a model-independent way with a reference cross section $\overline{\sigma}_e$, and a dark-matter form-factor $F_\text{DM}(q)$:
\begin{eqnarray}
\label{eq:sigma-bar_e}
\overline{\sigma}_e &\equiv& \frac{\mu_{\chi e}^2}{16 \pi m_\chi^2 m_e^2} \, \overline{\abs{\mathcal{M}_{\chi e} (q)}^2} \, \Big|_{q^2 = \alpha^2 m_e^2} \,,\\
\label{eq:DM-form-factor}
\overline{\abs{\mathcal{M}_{\chi e} (q)}^2} &=&
\overline{\abs{\mathcal{M}_{\chi e} (q)}^2} \, \Big|_{q^2 = \alpha^2 m_e^2}
\times \abs{F_\text{DM}(q)}^2\,.
\end{eqnarray}
$\overline{\sigma}_e$ is equal to the non-relativistic dark-matter--electron elastic scattering cross section, but with the 3-momentum transfer $q$ fixed to the reference value $\alpha m_e$ (appropriate for atomic processes). Here $\overline{\abs{\mathcal{M}_{\chi e} (q)}^2}$ is the squared matrix element for dark-matter--electron scattering, averaged over initial and summed over final spin states. We assume the DM form-factor has no directional dependence.

\mysubsection{Ionization in Atoms}
Dark matter may scatter with an electron bound in energy level $i$, ionizing it to an unbounded state with positive energy, $E_R = \frac{k'^2}{2m_e}$ (see also~\cite{Bernabei:2007gr,Kopp:2009et}). At large distances the unbound wavefunction $\tilde{\psi}_{k' l' m'}(\x)$ is that of a free spherical wave, but near the origin it is modified by the presence of the ion from which it escaped. Taking into account the density of unbound states, the thermally averaged differential cross section is given by:
\begin{equation}
\label{eq:ionization-cross section}
\!\!\! \frac{d \langle \sigma_{ion}^{i} v \rangle}{d\ln{E_R}} \!=\! \frac{\overline{\sigma}_e}{8 \mu_{\chi e}^2} \!\int\!\!\! q \, d q \big|f_{ion}^{i}(k',q)\big|^2 \big|F_\text{DM}(q)\big|^2 \eta (v_\text{\rm min}) \!
\end{equation}
where  $\eta(v_{\rm min}\!)$ has its usual meaning $\langle \frac{1}{v} \theta(v\!\!-\!\!v_{\rm min}\!) \rangle$ and $f_{ion}^{i}(k',q)$ is the form-factor for ionization:
\begin{equation}
\label{eq:ionization-form-factor}
\big|f_{ion}^{i}(k',q)\big|^2 \!=\! \frac{2 k'^3}{(2\pi)^3} \!\!\sum_{\substack{\text{degen.}\\ \text{states}}}\!\! \abs{\int\! d^3 x \, \tilde{\psi}^*_{k' l' m'}(\x) \psi_i (\x) e^{i \q \cdot\x}}^2
\end{equation}
Here the sum is over all final state angular variables $l'$ and $m'$, and over all degenerate, occupied initial states. The unbound wavefunctions are normalized to $\langle \tilde{\psi}_{k' l' m'}| \tilde{\psi}_{k l m}\rangle = (2\pi)^3 \delta_{l' l} \delta_{m' m} \frac{1}{k^2}\delta(k'-k)$.

In practice, since the correct unbounded wavefunctions are tedious to compute,
 it is useful to approximate the outgoing electron as a free plane wave.  In this case, for a spherically symmetric atom with full shells, the form-factor reduces to $(2l+1) k'^2/(4\pi^3 q) \int k\, d k \abs{\chi_{n l}(k)}^2$, with integration limits $\abs{k' \pm q}$. Here $\chi_{n l}$ is the radial part of the momentum-space wavefunction for the bound electron in the $n\, l$ shell, normalized to $\int k^2 d k \abs{\chi_{n l}(k)}^2 = (2\pi)^3$. The rate computed can then be corrected by the Fermi-factor, Eq.~(\ref{eq:fermifunc}), using an appropriate $Z_{\rm eff}$. We use the tabulated numerical RHF wavefunctions from~\cite{Bunge:1993}, and take $Z_{\rm eff}=1$, which is a slightly conservative choice since within the atom the true potential is somewhat larger.

As a cross-check, we also solved the radial Schr\"odinger equation for the exact unbound wavefunctions, using the effective potential extracted from the bounded wavefunctions directly, and computed the event rates according to Eq.~(\ref{eq:ionization-form-factor}). The rates calculated using the previous method agree with this more exact calculation to within $\OO(30\%)$ for outer-shell electrons, while for inner-shells, agreement requires somewhat larger $Z_{\rm eff}$. Since the outer-shell electrons dominate the total rate, this justifies our use of $Z_{\rm eff}=1$ in the Fermi-factor.

The above leads to a differential event rate,
\bea
\label{eq:excitation-rate}
\frac{dR_{ion}}{d\ln E_R}
& = & N_T \frac{\rho_\chi}{m_\chi} \frac{d\langle \sigma_{ion} v \rangle}{d\ln E_R}\\
& =& \frac{6.2}{A} \frac{\text{events}}{\text{kg-day}} \!
\bigg(\!\frac{\rho_\chi}{0.4 \frac{\text{GeV}}{\text{cm}^3}} \!\bigg) \!\!\! \left(\!\frac{\overline{\sigma}_e}{10^{-40}\text{cm}^2}\!\right) \!\!\! \nonumber \\
& & \times 
\left(\!\frac{10 \text{MeV}}{m_\chi}\!\right) \!\!
\frac{d\langle \sigma_{ion} v \rangle/d\ln E_R}{10^{-3} \overline{\sigma}_e}\,, \nonumber
\eea
where $N_T$ is the number of target nuclei per unit mass, $A$ is the mass-number of the target material, and $\rho_\chi$ is the local density of $\chi$.

\mysubsection{Ionizations in Crystals}
Due to their band structure, crystals have a great potential for significantly lowering the interaction threshold.   Upon scattering, an electron is excited from a valence band to a conduction band, where it may be drifted and detected.   The scattering rate is derived in a similar manner to that for excitations and ionizations.  The main difference lies in that the electrons reside in energy bands and are described via Bloch wave functions, $\psi_{i, \k}(\x)$,
\begin{gather}
\label{eq:blochWF}
\psi_{i,\k}(\x) = \frac{1}{\sqrt{V}}\sum_{G} \psi_i(\k+\G) e^{i(\k+\G)\x}\,.
\end{gather}
Here $i$ is the band index, $\k$ is the electron momentum in the first Brillouin Zone (BZ), $\G$ are the vectors in the reciprocal lattice and $V$ is the lattice volume.    

Since the crystal axis defines a preferred direction, the scattering rate depends in principle on the orientation of the crystal.  For an interaction that excites the electron from a valence energy band $i$  to a conduction band $i^\prime$, one finds the velocity averaged cross section,
\begin{eqnarray}
\label{eq:xsec2}
&&\langle\sigma_{cr}^{i\to i'} v\rangle = 
\\
&&\hspace{.8cm}\overline\sigma_e\int \frac{qdq}{\mu_{\chi e}^2}|F_{\rm DM}(q)|^2\int_{\rm BZ}\frac{Vd^3k}{(2\pi)^3} \langle {\cal F}^2_{i\rightarrow i'}({\q, \k,v_{\rm min}})\rangle\,,
\nonumber\\
&&\langle {\cal F}^2_{i\rightarrow i'}({\q, \k,v_{\rm min}})\rangle =
   \\
&&\hspace{.8cm}\int \frac{d^3v}{v} \int \frac{d\phi_v}{2\pi} |f_{cryst}^{i\rightarrow i'}({\q, \k})|^2  f_{\rm MB}(\v)\theta(v-v_{\rm min})\,,
\nonumber \\
&&v_{\rm min} = \frac{\Delta E_B}{q} +\frac{q}{2 m_\chi} \,.
\end{eqnarray}
Here, $f_{\rm MB}$ is the Maxwell-Boltzmann velocity distribution, and  $\phi_v$ is defined on the plane perpendicular to the direction of the incoming DM velocity, $v$.   
The form-factor, $f_{cryst}^{i\rightarrow i'}({\q, \k})$, is given by,
\begin{equation}
\label{eq:FFcrystal}
f_{cryst}^{i\rightarrow i'}({\q, \k}) = \sum_G \psi^*_{i'}(\k + \G+\q)\psi_i(\k + \G)\,.
\end{equation}
The energy gap is given by $\Delta E_B = E_{i'}(\k+\q) - E_i(\k)$ and hence the integrals over $\k$, $\v$, and $\phi_v$ all convolve the form-factor with the velocity distribution and encode the directional dependence of the rate.  
The study of directionality is postponed to future work.  
For simplicity, below we average over the form-factor, and take
\begin{equation}
\label{eq:FFaverage}
\langle {\cal F}^2_{i\rightarrow i'}({\q, \k,v_{\rm min}})\rangle = \int \frac{d\Omega}{4\pi} |f_{i\rightarrow i'}({\q, \k})|^2\eta(v_{\rm min})\,.
\end{equation}

We compute the crystal band structure and single electron wave functions using the {\sc Quantum ESPRESSO}~\cite{QE-2009} package which employs a local density approximation (LDA) within the density-functional theory.    The computation is done on a mesh of  k-vectors~\cite{Monkhorst:1976zz} and a regular grid of G-vectors with a cutoff, $|\k+\G|^2/2m_e<E_{\rm cut}$, taken to be $50\unit{Ry}$.  We use a BHS pseudopotential~\cite{BHS,PhysRevB.44.8503}, found in~\cite{QE-2009}.
The total cross section is obtained by summing over all occupied energy bands, $i$ and all conducting bands, $i^\prime$ in Eq.~\eqref{eq:xsec2}.  Hence  a large number of unoccupied states should, in principle, be included.   In practice, however, 
we find that including $24$ energy bands is sufficient, and corrections from higher bands are negligible.

\begin{figure}[t]
\begin{center}
\includegraphics[width=0.48\textwidth]{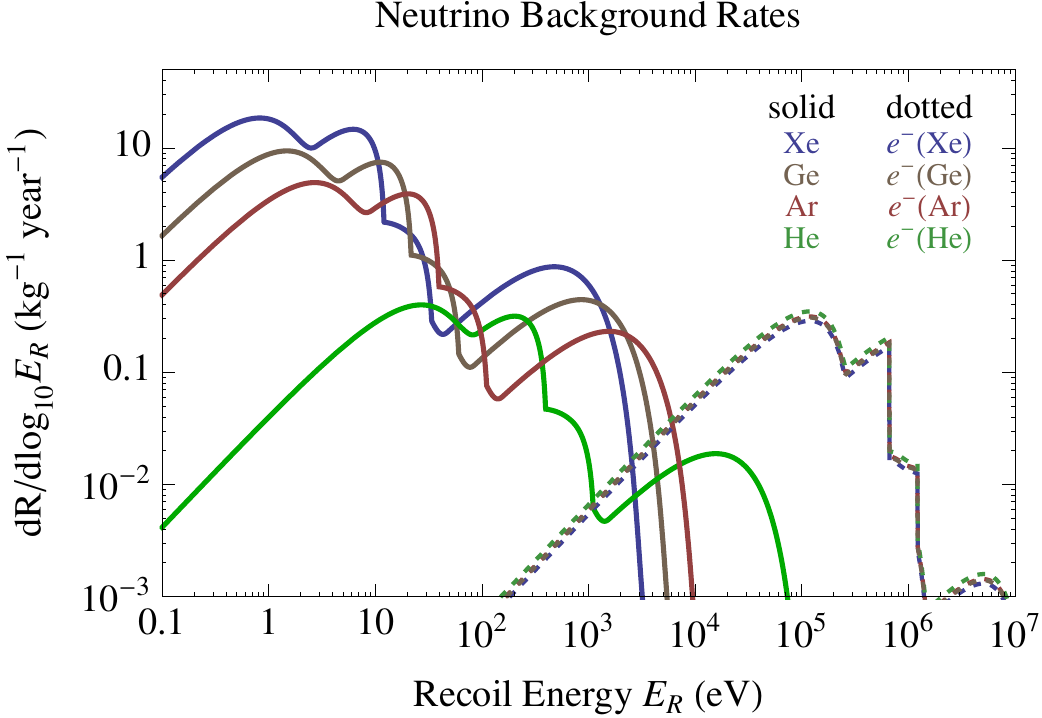}
\caption{Background solar neutrino rates per kg$\cdot$year.  Solid lines show  
\textit{nuclear} recoil spectra for neutrinos scattering with xenon (blue), germanium (brown), argon (red), 
and helium (green). 
These are not expected to significantly contribute to the ionized electron signal from LDM-electron scattering.  
Dotted lines, with same color coding as above, show rates for neutrino scattering off electrons.  
These rates are small and peak at higher energies than LDM-electron scattering.}
\label{fig:Neutrino Background}
\end{center}
\end{figure}

\begin{figure*}[t]
\begin{center}
\subfigure{
\includegraphics [width = 0.48\textwidth]{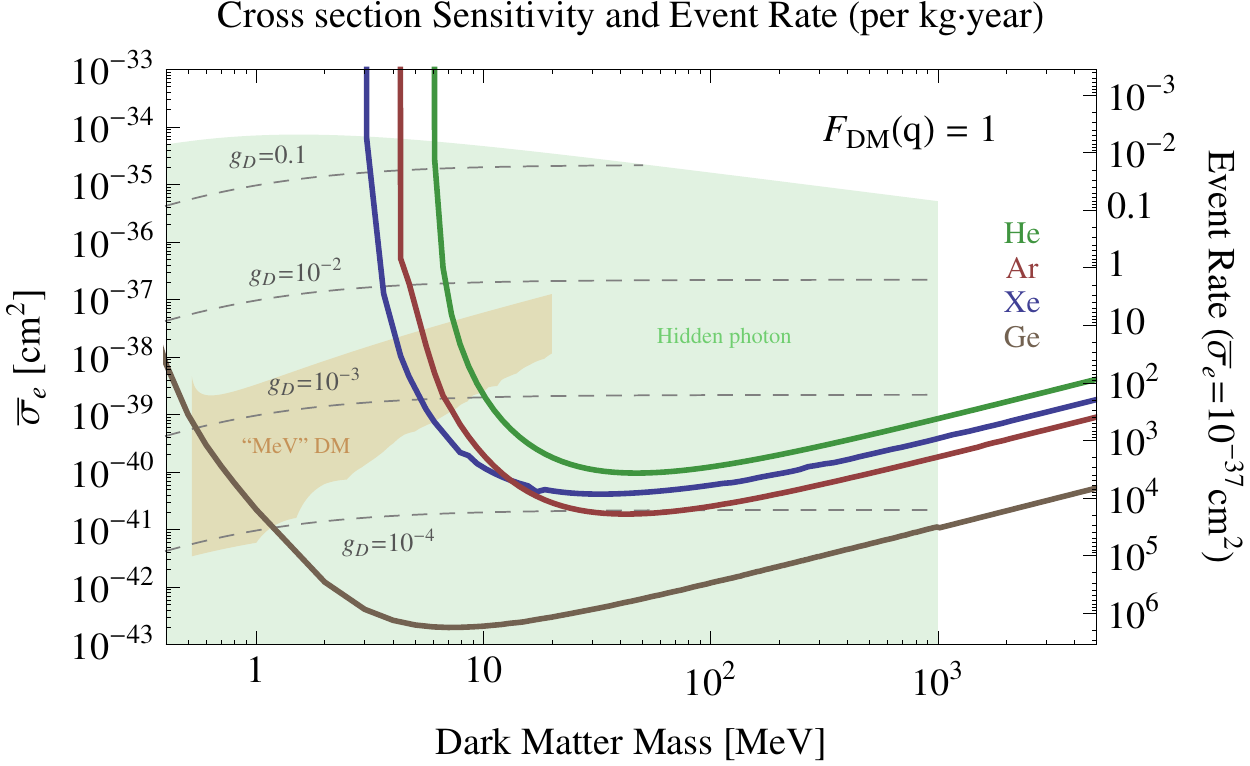}
}
\subfigure{
\includegraphics [width = 0.48\textwidth]{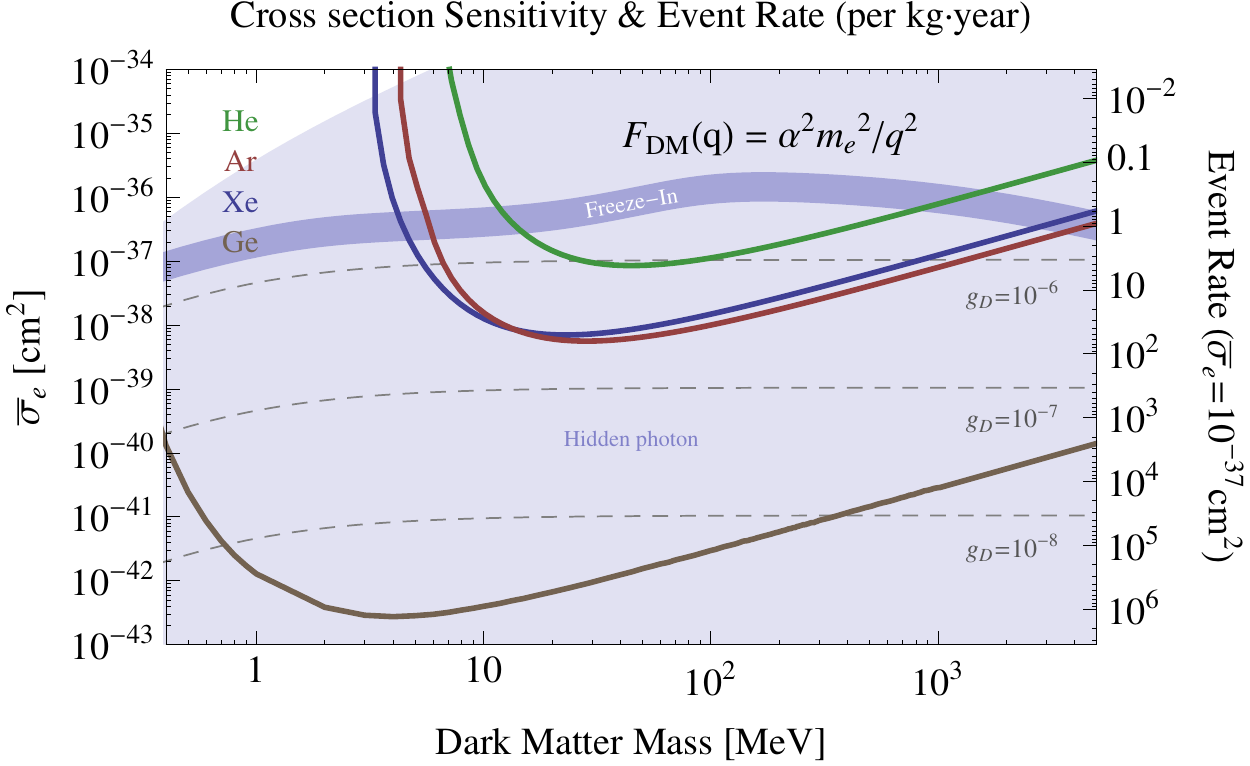}
}
\vskip -2mm
\caption{
The cross section exclusion reach (left axis) at 95\% confidence level for 1 kg$\cdot$year of exposure, assuming only the irreducible neutrino background (note that additional unknown backgrounds are likely to exist, which would weaken the 
sensitivity --- see Fig.~\ref{fig:sigmsVSbackground}). This corresponds to the cross section for which 3.6 events are expected after 1 kg$\cdot$year. The right axis shows the event rate assuming a cross section 
of $\overline{\sigma}_e=10^{-37}$ cm$^2$. Results are shown for xenon (blue), argon (red), germanium (brown), and helium (green) targets.
{\bf Left:} Models with no DM form-factor.  The green shaded area indicates the allowed region for $U(1)_D$ (hidden photon) models with $m_{\aprime} \simgt 10$ MeV. 
The orange shaded area is the region in which a particular model of ``MeV'' DM can explain 
the INTEGRAL 511 keV $\gamma$-rays from the galactic bulge \cite{Borodatchenkova:2005ct}. 
{\bf Right:} Models with a very light  scalar or vector mediator, for which $F_{\rm DM} = \alpha^2 m_e^2/q^2$. The blue region indicates the allowed parameter space for a hidden $U(1)_D$ model with a very light ($\ll$ keV) hidden photon.   
The darker blue band corresponds to the ``Freeze-In'' region.  
For illustration, constant $g_D$ contours are shown with dashed lines, assuming 
$m_{\aprime}=8\unit{MeV}$ and $\varepsilon=2\times10^{-3}$ (left plot) and 
$m_{\aprime}=1\unit{meV}$ and $\varepsilon=3\times10^{-6}$ (right plot).
For more details see the text and the Appendix.  
}
\label{fig:Sensitivity and Rates vs DM mass}
\end{center}
\end{figure*}

\section{BACKGROUNDS}
\label{sec:background}
Control over 
backgrounds is crucial for a successful LDM search. However, the backgrounds to very low energy signals, such as individual ionized electrons, are neither well measured nor well understood~\cite{Formaggio:2011jt, Sorensen:thesis, Edwards:2007nj, Angle:2011th}, and current direct detection experiments have not attempted to mitigate them. 
Although current technology is not able to distinguish individual LDM signal events from individual background events, 
one would expect that dedicated detector designs would allow significant improvements.  
Moreover, the annual modulation of the signal rate provides an additional handle to distinguish signal from background. 
Here we provide a brief qualitative discussion of several possible backgrounds, paying more attention to the well-understood and irreducible neutrino background.  

{\it Radioactive impurities}.  Radioactive decays typically deposit energy well above a keV, and so should be easily distinguished from the much lower energy DM signal. However, occasional low-energy events will occur, such as gamma rays escaping the detector after only a single, small-angle scatter, or electrons from the low-energy tail of beta-decay spectra. These events are phase-space suppressed by orders of magnitude relative to the total radioactive decay rate.

{\it Surface events}.  As in conventional direct detection experiments, higher-energy surface events may appear to have spuriously low energies due to partial signal collection. The position reconstruction required to reject this background may require new experimental designs, since existing detectors cannot reconstruct the $z$-position of very low energy events.

{\it Secondary events}.  
The primary signal of a higher-energy background may be accompanied by a number of very low energy events. This effect was observed for single-electron events in ZEPLIN-II~\cite{Edwards:2007nj} and \xenon~\cite{Sorensen:thesis, Angle:2011th}.  One possible explanation is 
the secondary ionization of impurities (e.g.~oxygen) or of xenon atoms by primary scintillation photons.  Such a background could be reduced by vetoing events occurring too close in time to a large event.  Another possible explanation is that electrons captured by impurities may eventually be released and detected a significant time after the primary event that produced them. The long lifetime of ionized impurities (e.g. an $O_2^-$ ion takes several seconds to drift to the anode in ZEPLIN-II) may limit the effectiveness of a timing veto, and in this case improvements in purification would be important.

{\it Neutrons}.  Current direct detection experiments are effective at shielding against neutron backgrounds. Modification of existing designs to minimize the very low energy neutron scattering relevant for 
LDM detection could yield further improvements.

{\it Neutrinos}. Neutrino scattering with electrons and nuclei generates a small but irreducible background. As with WIMP searches, this may set the ultimate limit to the reach of LDM direct detection experiments. 
The neutrino background is overwhelmingly dominated by solar neutrinos, which are theoretically well understood but only partially measured. Solar neutrinos have typical energies between $100\unit{keV}$ and $20\unit{MeV}$ and scatter with a rate given by:
\begin{equation}
\label{eq:NuRate}
\f{dR}{dE_R} = \int_{E_{\nu}^{\rm min}}^\infty \,dE_\nu \, \f{d\Phi_\nu}{dE_\nu}\, \f{d\sigma}{dE_R}\,, 
\end{equation}
where $E_{\nu}^{\rm min} \simeq \frac{1}{2}(E_R + \sqrt{E_R^2+2 E_R m} )$ is the minimal neutrino energy required to recoil a particle of mass $m$ with energy $E_R$, $d\sigma/dE_R$ is the scattering cross section, and $d\Phi_\nu/dE_\nu$ is  the solar neutrino flux~\cite{Bahcall:1997eg,Bahcall:1996qv,Bahcall:1987jc}. 
We calculate the differential rate for different materials in Fig.~\ref{fig:Neutrino Background} 
(see also e.g.~\cite{Freedman:1973yd,Monroe:2007xp,Strigari:2009bq}). Electron recoils have energies well above the expected DM signal and should be easily distinguished. Recoiling nuclei, on the other hand, have energies typically below a keV. The efficiency in converting this energy into ionized electrons is unknown at these low energies, but it is expected to be very small~\cite{Angle:2011th, Formaggio:2011jt}. Therefore the neutrino-induced background, for events in which only one or a few electrons are seen, is at most $\OO(1)$ per kg$\cdot$year and probably much lower.

\section{RESULTS}  
\label{sec:results}
%
%
We now present expected rates of ionization by DM--electron scattering in LDM direct detection experiments. A systematic study of possible target materials is beyond the scope of this letter, but we present illustrative results for xenon, argon, helium, and germanium. Noble gases and semiconductors, particularly xenon and germanium, respectively, are well established detector materials allowing internal amplification of ionized electrons by scintillation or phonon emission. As discussed, single electron sensitivity has already been achieved using xenon, while semiconductor targets benefit from low ionization thresholds (e.g., the bandgap in germanium is $0.7$ eV).

\begin{figure}[t]
\begin{center}
\subfigure{
\includegraphics [width = 0.48\textwidth]{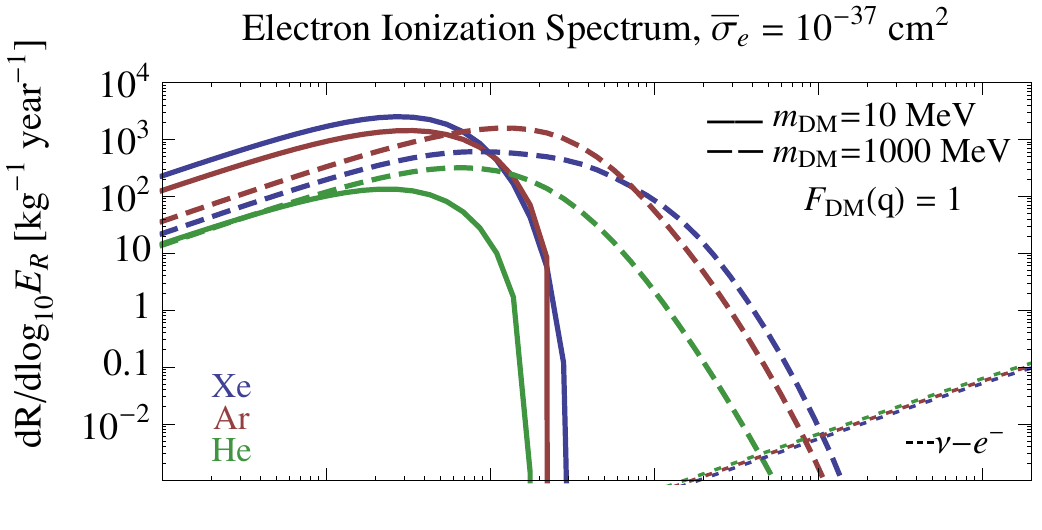}
}\\
\vskip -0.61cm
\subfigure{
\includegraphics [width = 0.48\textwidth]{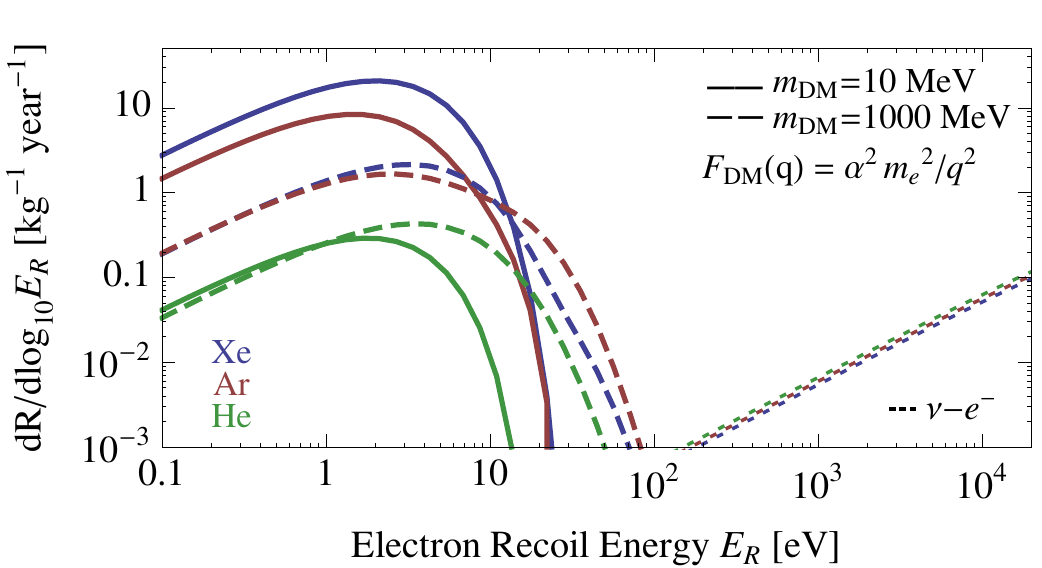}
}
\vskip -2mm
\caption{
The differential rates of LDM-induced ionization versus electron recoil energy, for a cross section 
of $\bar{\sigma}_{e} = 10^{-37}$~cm$^2$. Results are shown for xenon (blue), argon (red), and helium (green) targets, and a DM mass of 10 MeV (solid lines) and 1 GeV (dashed lines).
The two plots show results for scattering with  no DM form-factor ({\bf top}) and with $F_{\rm DM} = \alpha^2 m_e^2/q^2$ ({\bf bottom}). The dotted lines in the bottom right corner show the irreducible solar-neutrino--electron scattering backgrounds.
We emphasize that other backgrounds of an \emph{unknown} size can be expected at all energies, and will require a dedicated study to be measured and understood. 
}
\label{fig:ionizationHeavyDiff}
\end{center}
\end{figure}

Fig.~\ref{fig:Sensitivity and Rates vs DM mass} shows the expected 95\% exclusion reach after one kg$\cdot$year exposure for an experiment with only irreducible neutrino backgrounds (taken to be negligible with this exposure, as discussed). This corresponds to the cross section required to obtain 3.6 signal events~\cite{Feldman:1997qc}. Equivalently, the right axes give the event rate assuming a cross section of $\bar{\sigma}_{e} = 10^{-37}$ cm$^2$. The lines correspond to xenon (blue), argon (red), 
helium (green), and germanium (brown) targets, and the left and right plots are for models with a DM form-factor $F_{DM}=1$ and $F_{DM}=(\alpha m_e/q)^2$, respectively (cf. Eq.~\eqref{eq:DM-form-factor}). For small DM masses, the reach falls as the energy available approaches the ionization threshold. For larger DM masses, the cross section saturates, and the reach falls linearly with decreasing number density. It is clear that germanium's low ionization threshold gives it a significant advantage at low masses. It also allows it to probe smaller momentum transfer, which is beneficial for DM models with a $(\alpha m_e/q)^2$ form-factor. Here we take the DM halo to have a local density of $\rho\DM = 0.4$ GeV/cm$^3$, and a Maxwell-Boltzmann velocity distribution with mean velocity $v_0=220$ km/s and a hard cut-off at $v_{\rm esc}=650$ km/s.  We parametrize the Earth's velocity in the galactic frame as in \cite{Lewin:1995rx}.   Finally, we note that the results are shown assuming DM-electron interactions only.   When the DM is heavier than a few $100$'s of MeV, DM-nuclear interactions, if present, may also ionize electrons.  The small probability to do so may then be compensated by typically larger cross-sections. 

Our discussion so far has been model independent, but for concreteness we now discuss a simple and natural class of models, which could be probed by a LDM direct detection experiment.  
Consider a fermonic  DM particle, $\chi$, charged under a new Abelian gauge group $U(1)_D$ with gauge coupling $g_D$.  The $U(1)_D$ gauge boson $\aprime$
can obtain a small coupling $\varepsilon e$ 
to ordinary charged particles through kinetic mixing with the photon~\cite{Holdom:1985ag,Galison:1983pa}, mediating DM--electron scattering.  We parameterize the direct detection cross section as in Eqs~\eqref{eq:sigma-bar_e} and \eqref{eq:DM-form-factor}:
\bea\label{eq:model}
& & \overline{\sigma}_e = 
\frac{16 \pi \mu_{\chi e}^2 \alpha \alpha_D \varepsilon^2}{(m_{\aprime}^2+\alpha^2 m_e^2)^2}  
\simeq
\begin{cases}
\frac{16 \pi \mu_{\chi e}^2 \alpha \alpha_D \varepsilon^2}{m_{\aprime}^4}\,, & m_{\aprime} \gg \alpha m_e \\
\frac{16 \pi \mu_{\chi e}^2 \alpha \alpha_D \varepsilon^2}{(\alpha \, m_e)^4}\,, & m_{\aprime} \ll \alpha m_e
\end{cases}
\nonumber \\
\label{eq:FDM}
& & F_{DM}(q) = \f{m_{\aprime}^2+\alpha^2m_e^2}{m_{\aprime}^2+q^2} \simeq
\begin{cases}
1\,, & m_{A'} \gg \alpha m_e \\
\f{\alpha^2 m_e^2}{q^2}\,, & m_{\aprime} \ll \alpha m_e
\end{cases}
\eea
where $\alpha_D=g_D^2/4\pi$.  
Depending on the $\aprime$ mass, the DM form-factor $F_{\rm DM}$ is either constant or behaves as $1/q^2$.

In Fig.~\ref{fig:Sensitivity and Rates vs DM mass}, we show interesting regions for this class of models in the $m_\chi$--$\overline{\sigma}_e$ plane. The light green and blue regions in the left and right plots are the regions spanned by models satisfying all existing constraints, with $m_{\aprime}\gg \alpha m_e$ and  $m_{\aprime}\ll \alpha m_e$, respectively. The darker blue band in the right plot indicates the value of $\varepsilon$ for which the DM abundance is achieved by ``Freeze-In''~\cite{Hall:2009bx}.  For illustration, we also show constant $g_D$ contours with dashed lines, assuming $m_{\aprime}=8\unit{MeV}$ and $\varepsilon=2\times10^{-3}$ (left plot) and 
$m_{\aprime}=1\unit{meV}$ and $\varepsilon=7\times10^{-9}$ (right plot).  
The appendix below contains a brief discussion of how these regions are derived.  
Finally, we also show in Fig.~\ref{fig:Sensitivity and Rates vs DM mass} 
another viable LDM model.  
The orange region corresponds to a particular ``MeV" DM model 
(a Majorana fermion interacting with a $U$-boson from \cite{Borodatchenkova:2005ct}),  
which could explain the INTEGRAL 511 keV $\gamma$-rays from the 
galactic bulge \cite{Jean:2003ci} and remain consistent with Cosmic Microwave Background 
bounds~\cite{Padmanabhan:2005es,Slatyer:2009yq}.  

Although we do not attempt to calculate it here, it is important to consider how many electrons will be produced in a LDM scattering event. For example, in xenon a $30\unit{MeV}$ DM particle will typically ionize a 5p outer-shell electron (with binding energy $E_B=12.4$ eV), giving it insufficient recoil energy to ionize a second electron. However, for larger DM masses, the recoiling electron is increasingly likely to have enough energy to cause secondary ionizations. Heavier DM is also more likely to ionize a 5s ($E_B=25.7$ eV) or 4d ($E_B=75.6$ eV) shell electron, followed by the emission of a de-excitation photon which itself causes photoionization. In Fig.~\ref{fig:ionizationHeavyDiff}, we plot the differential ionization rate against electron recoil energy for xenon (blue), argon (red) and helium (green), for a DM mass of 10~MeV (solid lines) and 1~GeV (dashed lines). In germanium (not shown) the situation is complicated by the band structure but is qualitatively the same. Signal events in which more than one electron is collected could be crucial, firstly for experiments in which a single-electron threshold cannot be reached, and secondly since backgrounds to few-electron events may prove to be much smaller than for single-electron events.  For further details, see~\cite{XE10}. 
 
\begin{figure}[t]
\begin{center}
\includegraphics[width=0.48\textwidth]{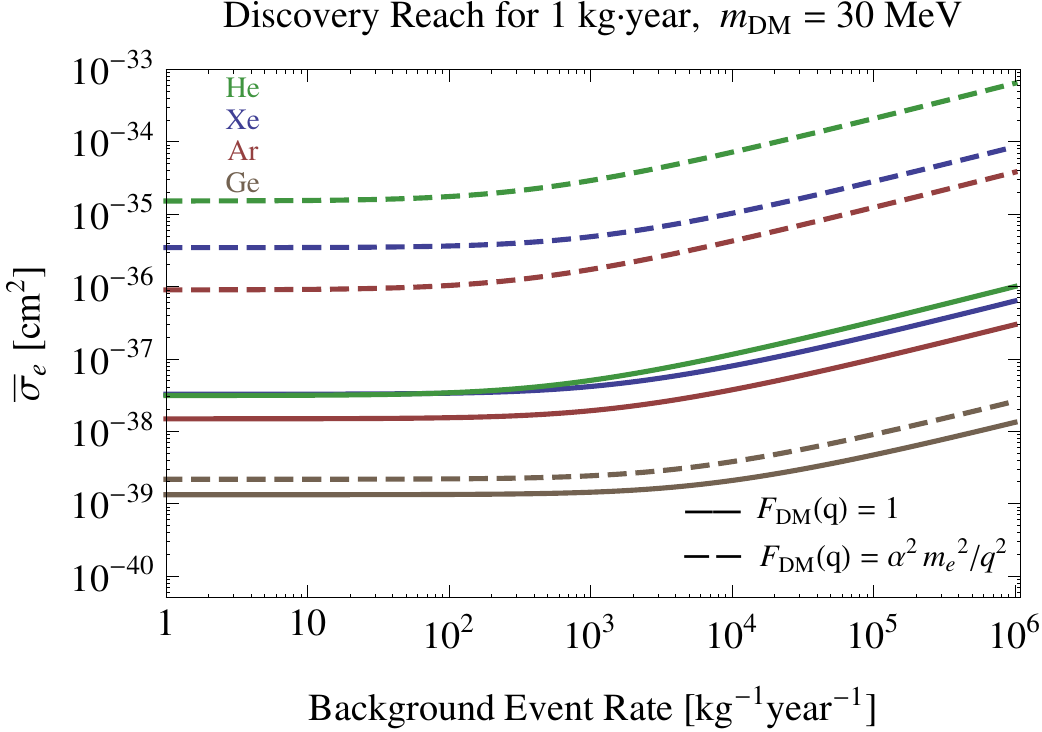}
\caption{
The discovery reach 
using annual modulation, as a function of the background event rate, for $m_{\rm DM} = 30$ MeV and 1 kg$\cdot$year exposure. Results are shown for xenon (blue), argon (red), germanium (brown) and helium (green) targets, assuming either no DM interaction form-factor (solid lines) or $F_{\rm DM} = \alpha^2 m_e^2/q^2$ (dashed lines). The annual modulation is $\OO(10\%)$ in all cases. The reach scales as $\sqrt{\text{exposure}}$ (exposure) for large (small) background rates. 
}
\label{fig:sigmsVSbackground}
\end{center}
\end{figure}
 
Besides neutrinos, the backgrounds to LDM scattering are currently largely unknown.  
An important handle to distinguish signal from background is therefore the annual modulation~\cite{Drukier:1986tm} of the DM scattering rate. Using the halo parameters given above, we find a modulation fraction $f_{\rm mod}$ of $\OO(10\%)$ for all cases considered, where $f_{\rm mod}$ is defined as the ratio of the modulating signal amplitude to the mean signal rate. A DM discovery would be possible by observing such a modulation over an unmodulated background. In Fig.~\ref{fig:sigmsVSbackground}, we show the modulation discovery reach as a function of the background event rate, for a DM mass of 30 MeV and for both constant (solid lines) and $(\alpha m_e/q)^2$ (dashed lines) DM form-factors.  Specifically, we calculate the cross section for which the modulated signal $\Delta S$ satisfies $\Delta S = 5 \sqrt{S_{\rm tot} + B}$, where $S_{\rm tot}$ is the total number of signal events, $B$ is the number of background events, and $\Delta S \equiv f_{\rm mod} S_{\rm tot}$.

As is clear from Figs.~\ref{fig:Sensitivity and Rates vs DM mass} and \ref{fig:ionizationHeavyDiff}, the rates can be very large for theoretically viable models.
This illustrates that there is a large discovery potential for the first experiments that attempt to explore this region.  
We encourage ongoing experiments such as XENON100, LUX, and CDMS, to 
actively pursue the required experimental sensitivity.

\subsection*{Acknowledgements}
We thank M.~Papucci for early collaboration. We also thank E.~Figueroa-Feliciano, R.~Harnik,  S.~Holland, J.~Kopp, A.~Malashevich, A.~Manalaysay,  P.~Meade, R.~Partridge, N.~Roe,  B.~Sadoulet,    
P.~Sorensen, H.~Yu, J.~Zupan, and K.~Zurek for many useful discussions.  We especially thank G.~Samsonidze for patiently helping us with {\sc Quantum ESPRESSO} and P.~Meade for significant computing support.
RE is supported by the US DOE under contract no.~DE-AC02-76SF00515. 
The work of JM and TV is supported in part by 
US DOE under contract DE-AC02- 05CH11231.
JM is also supported by NSF grants PHY-0457315 and PHY-0855653.
RE, JM, and TV acknowledge 
support from KITP and NSF Grant No. NSF PHY05-51164. RE and TV also acknowledge the hospitality of the Aspen Center for Physics and support from NSF Grant No.~1066293.

\appendix

\section{APPENDIX:~~MODEL CONSTRAINTS}\label{sec:app}

In the results section, we mentioned a simple class of models that can be probed by LDM 
direct detection experiments.  
We here briefly discuss the two interesting parameter regions that satisfy all existing constraints, 
leaving a more detailed discussion of the constraints and other LDM models to future 
work.  

We assume DM is charged under a new Abelian gauge group, $U(1)_D$, with gauge 
boson $\aprime$ and coupling $g_D$.  $\aprime$ couples with strength $\varepsilon e$  
to ordinary electrically charged particles via kinetic mixing with the 
hypercharge gauge boson \cite{Holdom:1985ag,Galison:1983pa}, and mediates DM--electron scattering.  
Theories with ``hidden'' sectors are natural and have recently received a lot of attention in 
other contexts, 
see e.g.~\cite{Strassler:2006im,ArkaniHamed:2008qn,Pospelov:2007mp,Pospelov:2008zw,Borodatchenkova:2005ct}.
Constraints on the $\aprime$-$\epsilon$ parameter space are reviewed in \cite{Jaeckel:2010ni}.  
Constrains on LDM coupling to an $\aprime$ have not been explored in detail, but can come from 
limits on, for example, DM annihilation-induced distortions of the Cosmic Microwave 
Background (CMB) signal and DM self-interaction induced distortions of DM halo 
shapes and the ``Bullet-Cluster'' dynamics.

Considering these constraints, two interesting parameter regions appear. 
The first is $m_{\aprime}\simgt 10\,\text{MeV}$, 
where there is a range of $\varepsilon$ in which the $\aprime$ 
is safe from beam-dump, collider, and muon- and electron-anomalous magnetic moment ($(g-2)_{\mu,e}$)   constraints~\cite{Bjorken:2009mm}, and may even resolve the discrepancy between the calculated and measured $(g-2)_{\mu}$~\cite{Pospelov:2008zw}. We note that the beam-dump and collider constraints are easily evaded if  the $\aprime$ decays to hidden sector rather than ordinary particles. 
Since $m_{\aprime}\gg \alpha m_e$, $F_{DM}=1$, see Eq.~\eqref{eq:FDM}. 
If $g_D$ is not too small, the visible and hidden sectors are thermalized in the early universe and $\chi$ 
for $m_\chi > m_{\aprime}$ efficiently annihilates to $\aprime$, making it natural to imagine an 
asymmetric DM abundance of $\chi$, with the asymmetry produced by new high-scale physics (see e.g.~\cite{Falkowski:2011xh}).  DM with $m_\chi<m_{\aprime}$ can instead annihilate to electrons through an 
off-shell $A'$ or to other hidden sector particles.  
$\aprime$ exchange generates DM self-interactions with cross-section 
$\sigma = g_D^4 m_\chi^2/4\pi m_{\aprime}^4$, which evades bounds from 
galaxy halo ellipticity \cite{Feng:2009hw} for  
$g_D \lesssim 0.1\,(\f{m_{\aprime}}{10~{\rm MeV}})\, (\f{m_\chi}{100~{\rm MeV}})^{-1/4}$
(for other similar or weaker self-interaction bounds 
see~\cite{MiraldaEscude:2000qt,Markevitch:2003at,Buckley:2009in}). 
Since DM is asymmetric, annihilation bounds from the CMB~\cite{Padmanabhan:2005es,Slatyer:2009yq},  which could otherwise be significant, do not apply. 
For $m_\chi > 1$~GeV, these models are constrained by the conventional direct detection limits of CRESST-I~\cite{Altmann:2001ax}. 
The allowed region in the $m_\chi - \overline{\sigma}_e$ plane is shown in green in Fig.~\ref{fig:Sensitivity and Rates vs DM mass}; $\overline{\sigma}_e$ is 
maximized for $m_{\aprime}\simeq 8$~MeV and $\varepsilon \simeq 2\times 10^{-3}$.

A second interesting region has a very light ($\ll \,\text{keV}$) or massless $\aprime$. Here 
DM--electron scattering has a form-factor $F_{DM}=(\alpha m_e/q)^2$.
For sufficiently small $g_D$, such models can then evade DM self-interaction 
bounds~\cite{,Feng:2009hw,Feng:2009mn}, numerous bounds on $\aprime$ kinetic mixing~\cite{Jaeckel:2010ni}, BBN bounds~\cite{Feng:2008mu,Falkowski:2011xh}, and (for $m_{\aprime}=0$) CMB bounds on millicharged DM~\cite{McDermott:2010pa}. The $\chi$ abundance can receive an irreducible ``Freeze-In''~\cite{Hall:2009bx} contribution from rare scattering processes in the early Universe thermal bath. Production of a hidden-photon population is highly suppressed by thermal effects~\cite{Jaeckel:2008fi, Redondo:2008ec}, but $\chi \overline{\chi}$ production occurs through 2$\to$2 s-channel annihilation of charged particles, and through decays of $Z$-bosons via the kinetic-mixing-induced coupling $\varepsilon g_D \tan{\theta_W} \overline{\chi} \gamma^\mu \chi Z_\mu$.
Our calculation of the former contribution agrees with~\cite{Chu:2011be}, but we also include the $Z$-decay contribution, which we find dominates for masses above $\OO$(GeV).
In the dark blue band in Fig.~\ref{fig:Sensitivity and Rates vs DM mass}, the DM abundance is entirely set by Freeze-In, 
an intriguing possibility consistent with the above bounds.
This fixes $\overline{\sigma}_e$ for a given $m_\chi$; e.g.~at $m_\chi=10$~MeV, 
$\varepsilon g_D \simeq 6\times 10^{-12}$ and $\overline{\sigma}_e\simeq 4\times 10^{-37}\,\text{cm}^2$.
Of course, other production mechanisms or annihilation to hidden-sector states may control the final DM abundance, and so a much larger range of parameters is possible. 
The light blue region in Fig.~\ref{fig:Sensitivity and Rates vs DM mass}
shows models satisfying both $\varepsilon < 3 \times 10^{-6}$, allowing $\OO$(\rm meV) hidden photons to evade various cosmological and laboratory bounds~\cite{Jaeckel:2008fi},
and the constraint from halo ellipticity,  
$g_D \lesssim 3\times 10^{-4}\,(\f{m_\chi}{100~{\rm MeV}})^{3/4}$ 
(there is a small logarithmic dependence on $m_{\aprime}$, which we set to 1 meV) \cite{Feng:2009hw}. 

\bibliography{lightdd}
\bibliographystyle{apsrev}

\end{document}